\documentclass[aps,prd,twocolumn,showpacs,preprintnumbers,amsmath,amssymb]{revtex4-1}

 \def\ep{{\epsilon}}

 \def\frac#1#2{{#1\over #2}}

 \def\s{\sqrt}

\def\be{\begin{equation}}
\def\ee{\end{equation}}
\def\ba{\begin{eqnarray}}
\def\ea{\end{eqnarray}}

 \def\de{\partial}

 \def\f {\frac}
 \def\ti{\tilde}
 \def\ap{\alpha}

 \def\no{\nonumber \\}

 \def\la{\langle}
 \def\lb{\rangle}
 \def\ep{\epsilon}

\usepackage{color}
\usepackage{graphicx}
\usepackage{dcolumn}
\usepackage{bm}

\begin{document}

\title{Holographic Dual of BCFT}
\author{Tadashi Takayanagi}
\affiliation{
Institute for the Physics and Mathematics of the Universe (IPMU),
University of Tokyo, Kashiwa, Chiba 277-8582, Japan
            }

\date{\today}

\begin{abstract}
We propose a holographic dual of a conformal field theory defined on
a manifold with boundaries, i.e. boundary conformal field theory
(BCFT). Our new holography, which may be called AdS/BCFT,
successfully calculates the boundary entropy or g-function in two
dimensional BCFTs and it agrees with the finite part of the
holographic entanglement entropy. Moreover, we can naturally derive
a holographic g-theorem. We also analyze the holographic dual of an
interval at finite temperature and show that there is a first order
phase transition.

\end{abstract}

\maketitle


{\bf{1. Introduction}}

\begin{figure}[bbb]
   \begin{center}
     \includegraphics[height=3cm]{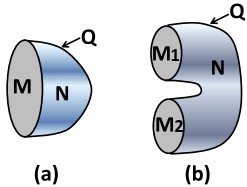}
   \end{center}
   \caption{Examples of the holographic duals of BCFT with a single AdS boundary (a)
   and two AdS boundaries (b).}\label{fig:setup}
\end{figure}

The AdS/CFT correspondence has been a very fascinating idea which
enables us to study quantum gravity in a non-perturbative way and at
the same time to analyze strongly coupled conformal field theories
(CFTs) efficiently \cite{Maldacena,GKP}. The purpose of this letter
is to consider the holographic dual of CFT defined on a manifold $M$
with a boundary $\de M$, which is so called boundary conformal field
theory (BCFT). We argue that this is given by generalizing the
AdS/CFT correspondence in the following way. Based on the idea of
holography \cite{holography}, we extend a $d$ dimensional manifold
$M$ to a $d+1$ dimensional asymptotically AdS space $N$ so that $\de
N=M\cup Q$, where $Q$ is a $d$ dimensional manifold which satisfies
$\de Q=\de M$. See Fig.\ref{fig:setup} for some examples of our
construction.

Usually, we impose the Dirichlet boundary condition on the metric at
the boundary of AdS and following this we assume the Dirichlet
boundary condition on $M$. On the other hand, we propose to require
a Neumann boundary condition on the metric at $Q$, whose details
will be explained later. This change of boundary condition is the most important
part of our holographic construction of BCFT. Our setup can be regarded as a modification 
of the well-known Randall-Sundrum setup \cite{RaSu} such that the additional boundary 
$Q$ intersects with the original asymptotically AdS boundary.
 See also \cite{CoMa} for an analysis of the Neumann boundary condition 
imposed at the asymptotically AdS boundary. Refer also to \cite{AMR}, where microscopic descriptions in string theory for a variety of boundary conditions in 
holographic setups have been discussed.\\

{\bf 2. Boundary Conditions}

To make the variational problem sensible, we usually add the
Gibbons-Hawking boundary term \cite{GHterm} to the Einstein-Hilbert
action (we omit the boundary term for $M$): \be I=\f{1}{16\pi
G_N}\int_{N}\s{-g}(R-2\Lambda)+\f{1}{8\pi G_N}\int_{Q}\s{-h}K.
\label{Loein} \ee The metric of $N$ and $Q$ are denoted by $g$ and
$h$, respectively. $K=h^{ab}K_{ab}$ is the trace of extrinsic
curvature $K_{ab}$ defined by $K_{ab}=\nabla_a n_b,$ where $n$ is
the unit vector normal to $Q$ with a projection of indices onto $Q$
from $N$.

Consider the variation of metric in the above action. After a partial integration, we find
\be
\delta I=\f{1}{16\pi G_N}\int_{Q}\s{-h}\left(K_{ab}\delta h^{ab}-Kh_{ab}\delta h^{ab}\right).
\ee
Notice that the terms which involve the derivative of $\delta h_{ab}$ cancels out
thanks to the boundary term.
We can add to (\ref{Loein}) the action $I_Q$ of some matter fields localized on $Q$.
We impose the Neumann boundary condition instead of the Dirichlet one by setting
the coefficients of $\delta h^{ab}$ to zero and finally
we obtain the boundary condition
\be
K_{ab}-h_{ab}K=8\pi G_N T^{Q}_{ab}, \label{bein}
\ee
where we defined
\be
T^{Qab}=\f{2}{\s{-h}}\f{\delta I_Q }{\delta h_{ab}}. \label{matbc}
\ee\\

{\bf 3. Construction of Holographic Dual of BCFT} As a simple
example we would like to assume that the boundary matter lagrangian
is just a constant. This leads us to consider the following action
\be I=\f{1}{16\pi G_N}\int_{N}\s{-g}(R-2\Lambda)+\f{1}{8\pi
G_N}\int_{Q}\s{-h}(K-T).\label{act} \ee The constant $T$ is
interpreted as the tension of the boundary surface $Q$. In AdS/CFT,
a $d+1$ dimensional AdS space (AdS$_{d+1}$) is dual to a $d$
dimensional CFT. The geometrical $SO(2,d)$ symmetry of AdS is
equivalent to the conformal symmetry of the CFT. When we put a $d-1$
dimensional boundary to a $d$ dimensional CFT such that the presence
of the boundary breaks $SO(2,d)$ into $SO(2,d-1)$, this is called a
boundary conformal field theory (BCFT) \cite{Cbcft}. Note that
though the holographic duals of defect or interface CFTs
\cite{KaRa,Janus} look very similar with respect to the symmetries,
their gravity duals are different from ours because they do not have
extra boundaries like $Q$.

To realize this structure of symmetries, we take the following
ansatz of the metric (see also \cite{KaRa,BE}): \be
ds^2=d\rho^2+\cosh^2\f{\rho}{R}\cdot ds^2_{AdS_{d}}.\label{metads}
\ee If we assume that $\rho$ takes all values from $-\infty$ to
$\infty$, then (\ref{metads}) is equivalent to the AdS$_{d+1}$. To
see this, let us assume the Poincare metric of AdS$_d$ by setting
\be ds_{AdS_d}^2=R^2\f{-dt^2+dy^2+d\vec{w}^2}{y^2},\label{metadss}
\ee where $\vec{w}\in R^{d-2}$. Remember that the cosmological
constant $\Lambda$ is related to the AdS radius $R$ by
$\Lambda=-\f{d(d-1)}{2R^2}$.

By defining new coordinates $z$ and $x$ as \be
z=y/\cosh\f{\rho}{R},\ \ x=y\tanh\f{\rho}{R}, \ee we recover the
familiar form of the Poincare metric of AdS$_{d+1}$:
$ds^2=R^2(dz^2-dt^2+dx^2+d\vec{w}^2)/z^2$.

\begin{figure}[ttt]
   \begin{center}
     \includegraphics[height=3cm]{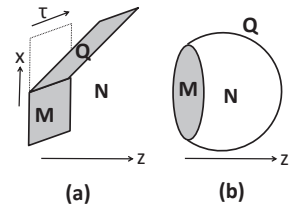}
   \end{center}
   \caption{The holographic dual of a half line (a) and a disk (b).}\label{fig:line}
\end{figure}

To realize a gravity dual of BCFT, we will put the boundary $Q$ at
$\rho=\rho_*$ and this means that we restrict the spacetime to the
region $-\infty<\rho<\rho_*$ (as described in
Fig.\ref{fig:line}(a)). The extrinsic curvature on $Q$ reads \be
K_{ab}=\f{1}{R}\tanh\left(\f{\rho}{R}\right)h_{ab} \ee The boundary
condition (\ref{bein}) leads to \be
K_{ab}=(K-T)h_{ab}.\label{eqbein} \ee Thus $\rho_*$ is determined by
the tension $T$ as follows \be
T=\f{d-1}{R}\tanh\f{\rho_*}{R}.\label{tension}
\ee\\

{\bf 4. AdS$_3/$CFT$_2$ and Boundary Entropy}

Let us concentrate on the $d=2$ case to describe the two dimensional
BCFT. This setup is special in that it has been well-studied (see
\cite{Cardy} and references therein) and that the BCFT has an
interesting quantity called the boundary entropy (or $g$-function)
\cite{AfLu}. The boundary state of a BCFT with a boundary condition
$\ap$ is denoted by $|B_\alpha\lb$ below. We define the quantity
called $g$ by the disk amplitude $g_\ap=\la 0|B_\alpha\lb$, where
$|0\lb$ is the vacuum state. The boundary entropy $S^{(\ap)}_{bdy}$
is defined by \be S^{(\ap)}_{bdy}=\log g_{\ap}. \label{gfun} \ee The
boundary entropy measures the boundary degrees of freedom and can be
regarded as a boundary analogue of the central charge $c$.


Consider a holographic dual of a CFT on a round disk defined by
$\tau^2+x^2\leq r^2_D$ in the Euclidean AdS$_3$ spacetime \be
ds^2=R^2\f{dz^2+d\tau^2+dx^2}{z^2},\label{poth} \ee where $\tau$ is
the Euclidean time. In the Euclidean formulation, the action
(\ref{act}) is now replaced by \be I_E=-\f{1}{16\pi
G_N}\int_{N}\s{g}(R-2\Lambda)-\f{1}{8\pi
G_N}\int_{Q}\s{h}(K-T).\label{acte} \ee Note that $\rho_*$ is
related to the tension $T$ of the boundary via (\ref{tension}). When
the BCFT is defined on the half space $x<0$, its gravity dual has
been found in previous section. Therefore we can find the gravity
dual of the BCFT on the round disk by applying the conformal map
(see e.g.\cite{BM}). The final answer is the following domain in
AdS$_3$ \be
\tau^2+x^2+\left(z-\sinh(\rho_*/R)r_D\right)^2-r_D^2\cosh^2(\rho_*/R)\leq
0. \label{diskh} \ee In this way we found that the holographic dual
of BCFT on a round disk is given by a part of the two dimensional
round sphere (see Fig.\ref{fig:line}(b)). A larger value of tension
corresponds to the larger radius.

Now we would like to calculate the disk partition function in order to obtain
the boundary entropy. By evaluating (\ref{acte}) in the domain (\ref{diskh}), we obtain
\ba
\!\!I_E\!=\!\f{R}{4G_N}\!\!\left(\!\f{r^2_D}{2\ep^2}\!+\!\f{r_D\sinh(\rho_*/R)}{\ep}\!+\!\log(\ep/r_D)\!-\!\f{1}{2}
\!-\!\f{\rho_*}{R}\!\!\right)\!\!,\ \ \ \ \label{diskpar}
\ea
where we introduced the UV cutoff $z>\ep$ as usual.
By adding the counter term on the AdS boundary \cite{Ren},
we can subtract the divergent terms in (\ref{diskpar}). The difference of the partition function
between $\rho=0$ and $\rho=\rho_*$ is given by $I_E(\rho_*)-I_E(0)=-\f{\rho_*}{4G_N}$.
Since the partition function is given by $Z=e^{-S_E}$, we obtain the boundary entropy
\be
S_{bdy}=\f{\rho_*}{4G_N}, \label{bed}
\ee
where we assumed $S_{bdy}=0$ for $T=0$ because the boundary contributions
vanish in this case.


Another way to extract the boundary entropy is to calculate the entanglement entropy.
The entanglement entropy $S_A$ with respect to the subsystem $A$ is
defined by the von Neumann entropy $S_A=-\mbox{Tr} \rho_A\log\rho_A$
for the reduced density matrix $\rho_A$. The reduced density matrix
is defined by tracing out the subsystem $B$, which is the complement
of $A$. In quantum field theories, we specify the subsystem $A$
by dividing a time slice into two regions. In a two dimensional CFT on a half line,
$S_A$ behaves as follows \cite{CaCa}
\be
S_A=\f{c}{6}\log\f{l}{\ep}+\log g,\label{entg}
\ee
where $c$ is the central charge and $\ep$ is the UV cut off (or lattice spacing);
$A$ is chosen to be an interval with length $l$ such that it ends at the boundary.
The $\log g$ in (\ref{entg}) coincides with the boundary entropy (\ref{gfun}).

In AdS/CFT, the holographic entanglement entropy is given in terms of the
 area of the codimension two minimal surface (called $\gamma_A$) which ends at $\de A$ \cite{RT}
\be S_A=\f{\mbox{Area}(\gamma_A)}{4G_N}. \ee Using this formula, the
boundary entropy in interface CFTs has successfully been calculated
in \cite{BE,BES}.

Consider the gravity dual of a two dimensional BCFT on a half line
$x<0$ in the coordinate (\ref{poth}). By taking the time slice
$\tau=0$, we define the subsystem $A$ by the interval $-l\leq x\leq
0$. In this case, the minimal surface (or geodesic line) $\gamma_A$
is given by  $x^2+z^2=L^2$. If we go back to the coordinate system
(\ref{metads}) and (\ref{metadss}), then $\gamma_A$ is simply given
by $\tau=0, y=l$ and $-\infty<\rho\leq\rho_*$. This leads to \be
S_A=\f{1}{4G_N}\int^{\rho_*}_{-\infty}d\rho. \ee By subtracting the
bulk contribution which is divergent as in (\ref{entg}),
we reproduce the previous result (\ref{bed}).\\

{\bf 5. Holographic g-theorem}

In two dimension, the central charge $c$  is the most important
quantity which characterizes the degrees of freedom of CFT.
Moreover, there is a well-known fact, so called c-theorem
\cite{cth}, that the central charge monotonically decreases under
the RG flow. In the case of BCFT, an analogous quantity is actually
known to be the g-function or equally boundary entropy \cite{AfLu}.
At fixed points of boundary RG flows, it is reduced to that of BCFT
introduced in (\ref{gfun}). It has been conjectured that the
g-function monotonically decreases under the boundary RG flow in
\cite{AfLu} and this has been proven in \cite{FrKo} later.
 Therefore the holographic proof of g-theorem described below
will offer us an important evidence of our proposed holography.
Refer to \cite{hcth} for a holographic c-theorem and to \cite{Ya}
for a holographic g-theorem in the defect CFT under a probe
approximation.

Because we want to keep the bulk conformal invariance and we know
that all solutions to the vacuum Einstein equation with $\Lambda<0$
are locally AdS$_3$, we expect that the bulk spacetime remains to be
AdS$_3$. We describe the boundary $Q$ by the curve $x=x(z)$ in the
metric (\ref{poth}). We assume generic matter fields on $Q$ and this
leads to the energy stress tensor $T^Q_{ab}$ term in the boundary
condition (\ref{bein}). It is easy to check the energy conservation
$\nabla^a T^{Q}_{ab}=0$ in our setup because
$\nabla^a(K_{ab}-Kh_{ab})=R_{n b}$, where $n$ is the Gaussian normal
coordinate which is normal to $Q$. In order to require that the
matter fields on the boundary are physically sensible, we impose the
null energy condition (or weaker energy condition) as in the
holographic c-theorem \cite{hcth}. It is given by the following
inequality for any null vector $N^a$ \be T^{Q}_{ab}N^aN^b\geq 0.
\label{nulle} \ee In our case, we can choose \be
(N^t,N^z,N^x)=\left(\pm
1,\f{1}{\s{1+(x'(z))^2}},\f{x'(z)}{\s{1+(x'(z))^2}}\right). \ee Then
the condition (\ref{nulle}) is equivalent to \be x''(z)\leq
0.\label{cond} \ee

Since at a fixed point the boundary entropy is given by
$S_{bdy}=\f{\rho_*}{4G_N}$ and we have the relation
$\f{x}{z}=\sinh(\rho_*/R)$ on the boundary $Q$, we would like to
propose the following $g$-function \be \log g(z)=\f{R}{4G_N}\cdot
\mbox{arcsinh}\left(\f{x(z)}{z}\right). \ee By taking derivative, we
get \be \f{\de \log g(z)}{\de z}=\f{x'(z)z-x(z)}{\s{z^2+x(z)^2}}.
\ee Indeed we can see that $x'z-x$ is non-positive because this is
vanishing at $z=0$ and (\ref{cond}) leads to $(x'z-x)'=x''z\leq 0$.
In this way, we manage to
derive the g-theorem in our setup.\\

{\bf 6. CFT$_2$ on Intervals and Phase Transitions}

Since so far we have studied a holographic BCFT in the presence of a
single boundary, next we would like to analyze a holographic dual of
a two dimensional CFT on an interval. At finite temperature, there
are two candidates for the bulk geometry, one of them is the thermal
AdS$_3$ and the other is the BTZ black hole (AdS$_3$ black hole). In
the absence of boundaries $Q$, there is the well-known Hawking-Page
phase transition between them \cite{HP,Wi}.


At low temperature, the bulk geometry is expected to be given by the thermal AdS$_3$ defined by
the metric
\be
ds^2=R^2\f{d\tau^2}{z^2}+R^2\f{dz^2}{h(z)z^2}+\f{R^2h(z)}{z^2}dx^2, \label{tads}
\ee
where $h(z)=1-(z/z_0)^2$. The periodicity of the Euclidean time $\tau$, denoted by
the inverse temperature $1/T_{BCFT}(\equiv 2\pi z_H)$,
can be chosen arbitrary, while that of the space direction
$x$ is determined to be $2\pi z_0$ by requiring the smoothness.

We again describe the boundary $Q$ by the curve $x=x(z)$. The boundary condition (\ref{eqbein})
is solved as follows
\be
x(z)-x(0)=z_0\cdot \arctan\left(\f{RTz}{z_0\s{h(z)-R^2T^2}}\right).  \label{thsx}
\ee

\begin{figure}[ttt]
   \begin{center}
     \includegraphics[height=3cm]{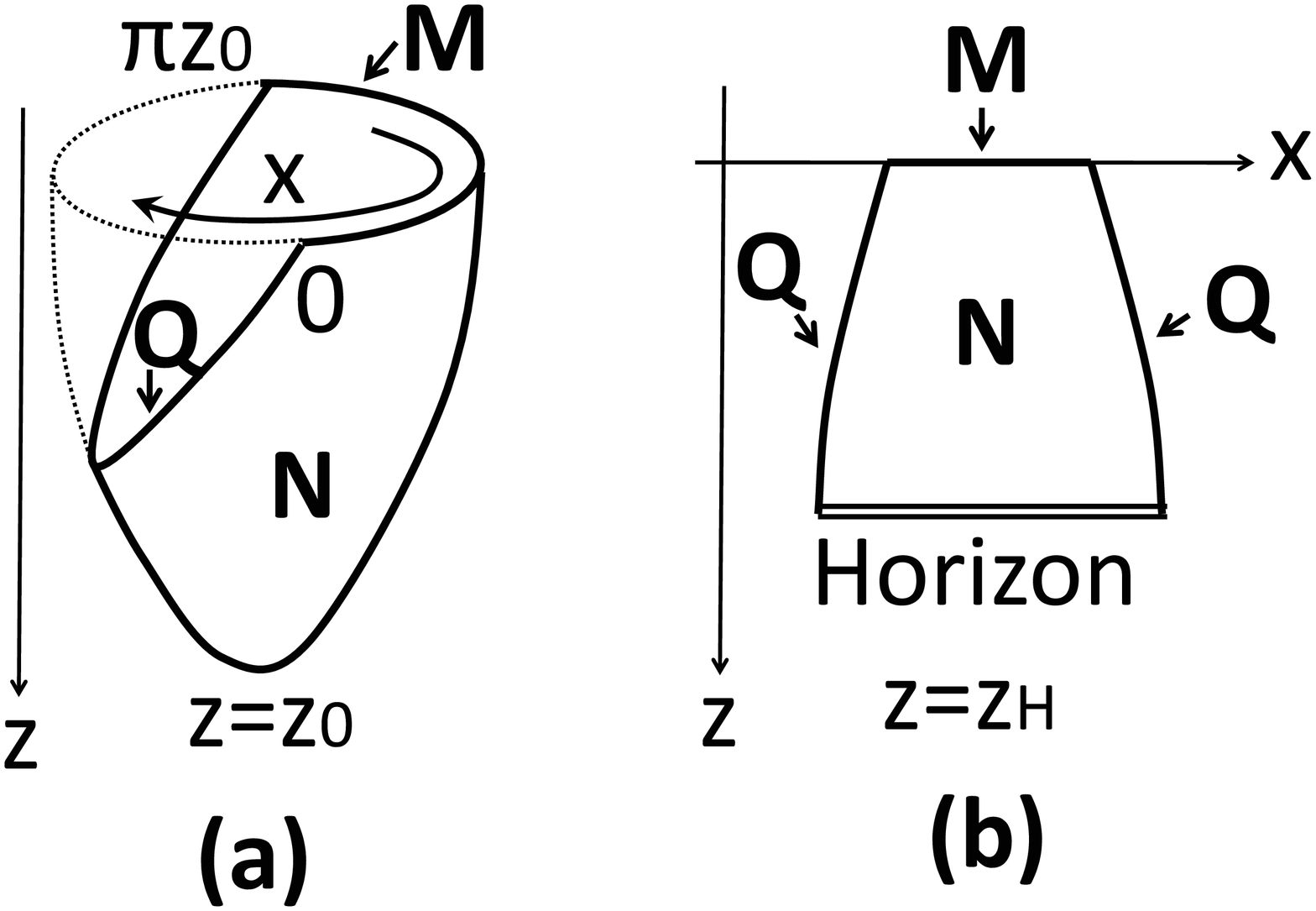}
   \end{center}
   \caption{The holographic dual of an interval at low temperature
   (a) and high temperature (b).}\label{fig:transition}
\end{figure}

Notice that $x'(z)$ gets divergent at $z_*=z_0\s{1-R^2T^2}$
and thus this should be the turning point (see Fig.\ref{fig:transition}(a)).
Thus totally the boundary $Q$ extends from $x=0$ to $x=\pi z_0$. Assuming $T>0$,
the bulk spacetime $N$ is defined by the sum of
$(-\pi z_0\leq x\leq 0,\  0<z\leq z_0) $ and $(0<x\leq \pi z_0,\  z(x)<z<z_0)$, where
$z(x)$ is the inverse function of (\ref{thsx}) and its extension to $\f{\pi}{2}z_0<x<\pi z_0$.

Now the Euclidean action (\ref{acte}) reads \ba
&&I_E=\f{Rz_H}{G_N}\left[\int^{z_*}_\ep\f{dz}{z^3}\left(x(z)+\f{\pi
z_0}{2}\right)+\int^{z_0}_{z_*} \f{dz}{z^3}(\pi z_0)\right]\no &&\ \
\ \ -\f{z_H TR^2}{2G_N}\int^{z_*}_\ep \f{dz}{z^2\s{h(z)-R^2T^2}},
\label{tadsac} \ea where $\ep$ is the UV cut off as before. To
evaluate (\ref{tadsac}) by eliminating the divergence, we need to be
careful in that we have to regard $2\pi \ti{z_0}$ as the physical
radius, defined by $\ti{z_0}=\s{f(\ep)}z_0$, matching the asymptotic
geometry at $z=\ep$. Also the contribution Gibbons-Hawking term at
the AdS boundary $M$ is vanishing as usual, by using the boundary
integral of $K-K^{(0)}$ instead of that of $K$, where $K^{(0)}$ is
the trace of extrinsic curvature for the pure AdS$_3$ (\ref{poth}).
In the end, we obtain the result \be I_E=-\f{\pi
Rz_H}{8G_Nz_0}=-\f{\pi}{24}\cdot\f{c}{\Delta x\cdot T_{BCFT}},
\label{tadsi} \ee where we employed the well-known relation between
the AdS$_3$ radius $R$ and the central charge $c$ of CFT$_2$, given
by $c=\f{3R}{2G_N}$ \cite{BrHe}. Note that the final result
(\ref{tadsi}) does not depend on the tension $T$ and is correct even
when $T<0$.


On the other hand in the higher temperature phase, the bulk is
described by a part of the BTZ black hole
\be
ds^2=R^2\f{f(z)}{z^2}d\tau^2+R^2\f{dz^2}{f(z)z^2}+R^2\f{dx^2}{z^2},
\ee
where $f(z)=1-(z/z_H)^2$. The Euclidean time $\tau$ is compactified on a circle such that
$\tau\sim \tau + 2\pi z_H$ and thus the temperature in the dual BCFT is $T_{BCFT}=\f{1}{2\pi z_H}$.
The length of the interval is again denoted by
$\Delta x=\pi z_0$.

We find the following profile $x=x(z)$ of $Q$ from (\ref{eqbein})
\be x(z)-x(0)=z_H\cdot
\mbox{arcsinh}\left(\f{RTz}{z_H\s{1-R^2T^2}}\right).  \label{solx}
\ee Note $Q$ consists of two disconnected parts as in
Fig.\ref{fig:transition}(b).

Now we evaluate the Euclidean action (\ref{acte}) in the form
$I_E=2I_{bdy}+I_{bulk}$. $2I_{bdy}$ is the boundary contributions,
while $I_{bulk}$ is the bulk ones which do not depend on $T$. After
subtracting the divergences, we obtain \be I_{bulk}=-\f{\pi
c}{6}\Delta x \cdot T_{BCFT}. \label{bulkbtz} \ee This result
(\ref{bulkbtz}) clearly agrees with what we expect from the standard
CFT results. On the other hand, each of two boundary contributions
is found to be \be
I_{bdy}=-\f{\rho_*}{4G_N}=-\f{c}{6}\mbox{arctanh}(RT).
\label{bdybtz} \ee The total thermal entropy of this thermal system
is found from (\ref{bulkbtz}) and (\ref{bdybtz}) \be
S_{thermal}=\f{\pi}{3}c\Delta x\cdot
T_{BCFT}+\f{c}{3}\mbox{arctanh}(RT) \ee

This calculation offers us one more independent calculation of
boundary entropy $S_{bdy}$ in AdS/CFT. Consider a BCFT at a finite
temperature $T_{BCFT}$, in other words, a CFT defined on a cylinder.
The two boundary conditions imposed on the two boundaries are
denoted by $\ap$ and $\beta$. They are described by the boundary
states $|B_{\ap}\lb$ and $|B_{\beta}\lb$. The partition function
$Z_{\ap\beta}$ on a cylinder, whose length is denoted by $\Delta x$,
gets factorized in the high temperature limit $T_{CFT}\Delta x>>1$
\be \la B_{\ap}|e^{-H\Delta x}|B_{\beta}\lb\simeq g_\ap g_\beta
e^{-E_0 \Delta x}, \label{bstate} \ee where $H$ is the Hamiltonian
(in the closed string channel) and $E_0$ is the ground state energy.
The final factor $e^{-E_0\Delta x}$ is interpreted as the thermal
energy for the CFT as is clear in the open string channel. Therefore
the contribution from the presence of boundary is the product of
g-function $g_\ap g_\beta$ \cite{AfLu}. In our holographic
calculation, this means $g=e^{S_{bdy}}=e^{-I_{bdy}}$ and this is
indeed true by comparing (\ref{bdybtz}) and (\ref{bed}).


Let us examine when either of the two phases is favored.
To see this we compare (\ref{bulkbtz})$+2\times$(\ref{bdybtz})
with (\ref{tadsi}) and pick up the smaller one. In this way
we find that the black hole phase is
realized when
\be
\Delta x\cdot T_{BCFT}>-\f{1}{\pi}\mbox{arctanh}(RT)+\s{\f{1}{4}
+\f{1}{\pi^2}\mbox{arctanh}^2(RT)}.\nonumber
\ee
At lower temperature, the thermal AdS phase is favored.
At vanishing tension $T=0$, the phase boundary $z_0=z_H$ coincides
with that of the Hawking-Page transition \cite{Wi}. As the tension
gets larger, the critical temperature gets lower. This is consistent
with the fact that the entropy
$S_{bdy}$ carried by the boundary increases as the tension does.
This phase transition is first order and is analogous to the
confinement/deconfinement transition in gauge theories \cite{Wi}.\\

{\bf 7. Conclusions and Discussions}
In this letter we proposed a holographic dual of BCFT. The crucial idea which extends the standard
AdS/CFT to our AdS/BCFT is to consider not only Dirichlet but also
Neumann boundary condition of the metric
at the same time. This clearly opens up a new stage of holography.
For example, it is interesting to consider the case where the boundary
$M$ consists of two disconnected manifolds $M_A$ and $M_B$
as in Fig.\ref{fig:setup} (b). The holographic entanglement
entropy $S_A$ between $M_A$ and $M_B$ is estimated as the minimal
area of the cross section of the throat \cite{HRT}, which is finite
and non-vanishing. Therefore this `open wormhole' geometry, if exists,
seems to argue that $M_A$ and $M_B$ are entangled, though disconnected.
Many things are left for future works such as the studies of correlation functions,
higher dimensional and
supersymmetric examples,
string/M theory realizations and applications to condensed matter physics.\\

{\bf Acknowledgments} TT would like to thank M.~Fujita,
S.~Mukohyama, N.~Ogawa, E.~Tonni, T.~Ugajin for discussions, and
especially J.~McGreevy and S.~Ryu for useful comments. TT is very
grateful to the Aspen center for physics and the Aspen workshop
``Quantum Information in Quantum Gravity and Condensed Matter
Physics,'' where this work was completed. TT is supported by World
Premier International Research Center Initiative (WPI Initiative),
MEXT, Japan. TT is supported in part by JSPS Grant-in-Aid for
Scientific Research No.\ 20740132, and by JSPS Grant-in-Aid for
Creative Scientific Research No.\ 19GS0219.

\end{document}